\documentclass[twocolumn,superscriptaddress,showpacs]{revtex4}
\usepackage{graphicx}
\def\hmpc{h^{-1}{\rm Mpc}}
\def\lsim{~\rlap{$<$}{\lower 1.0ex\hbox{$\sim$}}}
\def\bsim{~\rlap{$>$}{\lower 1.0ex\hbox{$\sim$}}}

\def\kms{\ {\rm Km\,s^{-1}}}
\def\hmpc{\ {\rm h^{-1}Mpc}}

\def\dd{{\rm d}}
\def\ln{{\rm ln}}
\def\pa{\partial}

\def\grad{\nabla}
\def\pmb#1{\setbox0=\hbox{#1}%
\kern-.025em\copy0\kern-\wd0
\kern.05em\copy0\kern-\wd0
\kern-.025em\raise.0433em\box0}

\def\ph{\varphi}

\def\phg{\ph_g}

\def\vv{\pmb{$v$}}

\def\vg{\pmb{$g$}}

\def\vx{\pmb{$x$}}

\def\vr{\pmb{$r$}}

\def\vC{\pmb{$C$}}

\def\phig{\phi_g}

\def\da {{\dot a}}

\def\qn{{q_{_n}}}

\def\vnabla{\pmb{$\nabla$}}

\usepackage{bm}

\begin{document}

\title{Boundary-value problems  in cosmological dynamics}

% First author block:
\author{Adi Nusser}
\email{adi@physics.technion.ac.il}
% \homepage{An author's web page; optional}
\affiliation{Physics Department- Technion, Haifa 32000, Israel}
% You may list several affiliation, using separate commands for each:
%\affiliation{A second affiliation (optional)}
%\affiliation{The third affiliation is shared by both co-authors}

% For other authors please repeat the author block as needed
%\author{Second Author}
% Note how REVTeX 4 deals with identical affiliations
%\affiliation{The third affiliation is shared by both co-authors}

\begin{abstract}

The dynamics of cosmological gravitating system is governed by the Euler and the Poisson equations. 
Tiny fluctuations near the big bang singularity are amplified by 
gravitational instability into the observed structure today. 
Given the current distribution of galaxies and 
assuming initial homogeneity, dynamical reconstruction methods 
have been developed to derive  the cosmic density and velocity fields back in time.  
The reconstruction method described here is based a least action principle formulation of 
the dynamics of collisionless particle (representing galaxies).
Two observational data sets will be considered. 
The first is the distribution of galaxies which is assumed 
to be an honest tracer of the mass density field of the dark matter. 
The second set is measurements of the peculiar velocities (deviations 
from pure Hubble flow) of galaxies.
Given the first data set, the reconstruction method recovers the associated velocity 
 field which can then be compared with the
second data set.
This comparison  constrains the nature of the dark matter and the 
relation between mass and light in the Universe.

\end{abstract}

% Insert suggested PACS numbers (up to 4) in braces.
% The PACS (Physics and Astronomy Classification Scheme) 
% can be accessed on the web at http://www.aip.org/pacs/

% Insert keywords (up to about 4) in braces; optional.
% \keywords{Up to four keywords}

\maketitle

% Here the text of your article begins

\section{\label{intro}Introduction}
% References should be done using the \cite, \ref, and \label commands.
% Put \label in argument of \section for cross-referencing like this:
%\section{\label{}}

Cosmology is concerned with observing 
and modeling the universe on large scales: from our own Milky Way, other 
galaxies, galaxy clusters, super clusters up to the largest scales as probed by measurements 
of the cosmic microwave background radiation (CMB).
These observations span a huge  range of scales and all strongly suggest that:
1) the dominant form of matter is dark (a factor of 6 in mass over the normal baryonic matter), 2) the clustering amplitude decreases with scale, and 
3) structure forms by gravitational amplification of ting initial fluctuations.
These are some of the main component of the standard paradigm in cosmology. 
Violation of any of them or all of them is  
consistent  with only a very limited set of observations, if any.  
Cosmology has had 
a great impact on other fields of physics and science in general.
The shear existence of the  gravitationally dominant dark matter has stimulated
scientists' (and others')  vivid imagination for a few decades now. 
Abundance and masses of non-standard particles 
have been constrained from the observed clustering pattern alone. 
In addition to gravity,
hydrodynamical processes can greatly influence the formation and
evolution of galaxies, groups and clusters of galaxies.
Hydrodynamical effects, however, play a minor role in shaping the
observed distribution of galaxies on scales a few times larger than
the size of galaxy clusters. Therefore, gravitational instability
theory directly relates the present-day large scale structure to the
initial density field and provides the frame-work within which the
observations are analyzed and interpreted.  Gravitational instability
is a non-linear process.  Analytic solutions exist only for
configurations with special symmetry, and approximate tools are
limited to moderate density contrasts.  So, numerical methods are necessary
for a full understanding of the observed large scale
structure of the universe.  There are two complementary numerical approaches.  The
first approach relies on $N$-body techniques designed to solve an
initial value problem in which the evolution of a self-gravitating
system of massive particles is determined by numerical integration of
the Newtonian differential equations.  Combined with semi-analytic
models of galaxy formation,  $N$-body simulations have become an
essential tool for comparing the predictions of cosmological
models with the observed properties of galaxies.  Because the exact
initial conditions are unknown, comparisons between simulations and
observations are mainly concerned with general statistical properties.
The second approach aims at finding the past orbits of mass tracers
(galaxies) from their observed present-day distribution. 
The orbits must
be such that the initial spatial distribution is homogeneous.  This approach
is very useful for direct comparisons between different types of
observations of the large scale structure.  Most common are the
velocity-velocity (hereafter {\it v-v}) comparisons between the
observed peculiar velocities of galaxies and and the velocity field
inferred from the galaxy distribution in redshift surveys.  This
types of analysis yield the cosmological mass density parameter
$\Omega_m$.
Any systematic mismatch between the fields 
serves as an indication to the nature of
galaxy formation and/or the origin of galaxy intrinsic scaling
relations used to measure the distances, provided that errors
in the calibration have been properly corrected for.  
This second approach also allows to
perform back-in-time reconstructions of the density field on scales
$\sim 5 \hmpc$  \cite{Frisch2002}.

Finding the orbits that satisfy initial homogeneity and match the
present-day distribution of mass tracers is a boundary value
problem. This problem naturally lends itself to an application of
Hamilton's variational principle where the orbits of the objects are
found by searching for stationary variations of the action subject to
the boundary conditions.  The use of the Principle of Least Action in
a cosmological frame-work has been pioneered by Peebles (1989) \cite{Peebles1989} and 
has long been restricted to small systems such as the Local Group \cite{Peebles1994}
and the Local Supercluster \cite{Shaya}. Early
applications to large galaxy redshift surveys have been hampered by
the computational cost of handling the relatively large number of
objects.  Subsequent numerical applications speeded
up the method and allowed the reconstruction of the orbits of $\sim
10^3$ particles (Shaya, Peebles \& Tully 1995).  However, it was only
recently that the improvement of the minimization techniques and the
use of efficient gravity solvers made it possible to deal with more
than $10^4$ objects\cite{NusserBranchini}, comparable to the number of objects
contained in the largest all-sky galaxy catalogs.

\section{Cosmological Dynamics}
   For the background cosmology we work with a Friedmann-Robertson-Walker Universe. In this uniform background, the physical distance, $r$, between two points is $r\propto a(t)$
   where $a(t)$ is the scale factor.
We consider a matter dominated universe with mean density 
$\bar \rho=\Omega \rho_c$ with $\rho_c=3H^2/8\pi G$. For a $\Omega=1$, we get a 
critical density flat universe with $a\sim t^{2/3}$. The Universe is geometrically open for $\Omega<1$ and close for $\Omega>1$.
Current observations indicate that the Universe contain a cosmological constant which makes it flat even though $\Omega\approx 0.3$\cite{Spergel}. Apart from the dependence of $a$ on $t$ the presence of 
a cosmological constant has very little effect on our description here. In particular, the equations of motion of perturbations remain correct. 
We further define,   $H(t)=\da /a$ is the Hubble function, and  denote the comoving coordinate of a patch of matter by $\vx={\vr }/a $. The 
fluctuations are described by the density contrast $\delta(\vx,t)=\rho(\vx,t)/\bar \rho(t)-1$ and 
the  comoving velocity by $\vv=\dd \vx /\dd t$.  
Also,
let $D(t)$ be the linear density growing mode normalized to unity at the
present epoch, and $f(\Omega_m)=\dd \ln D/\dd \ln a \approx
\Omega_m^{0.6}$ (e.g., Peebles 1980).  
The equations governing the evolution of fluctuations 
in a collisionless mass component in an expanding Universe are, 
The Euler equation,
\begin{equation}
\frac{\dd \vv}{\dd t} +2 H \vv +\vv \cdot \vnabla \vv =
-\vnabla \phg , 
\label{eq:euler}
\end{equation}
the continuity, 
\begin{equation}
\frac{\pa \delta}{\pa t}+\vnabla \cdot  (1+\delta) \vv =0
\end{equation}
and the Poisson equation,
\begin{equation}
\grad^2 \phg=4\pi G \bar \rho \delta \; .
\end{equation}
The term $2H\vv$ in the Euler equation is due  to the 
expansion of the cosmological background. 
The source term in the Poisson equation represents density fluctuations above the 
mean background density. 

\subsection{Linear gravitational instability}
Neglecting the non-linear terms $\vv \cdot \vnabla\vv$ and 
$\vnabla \cdot \delta \vv$, 
 the equations of motion reduce to 
 \begin{equation}
 \delta=-\frac{1}{f(\Omega) H}\vnabla \cdot \vv \; 
\label{eq:linv}
\end{equation}
and
\begin{equation}
\ddot \delta+ 2H \dot \delta=\frac{3}{2}H^2\Omega\delta \; ,
\label{eq:lind}
\end{equation}
where an over-dot indicates a time derivative. For a critical density 
Universe ($\Omega=1$  and $H=2/3t$), the equation (\ref{eq:lind})  gives 
$\delta_1 \propto t^{2/3} $ and $\delta_2\propto t^{-1}$, as the growing and 
decaying solutions, respectively. 
A few  things to note. First, without the term $ 2H\dot \delta$ the
solutions would be exponential functions rather than power laws in time. 
Second, even in the linear regime, the decaying mode
prevents a full recovery of the initial conditions, at $t\approx 0$ near the big
bang cosmological singularity. Indeed, recovering this 
mode requires a precise knowledge of the present  $\delta$ and $\dot \delta $ (or $\vv$), in order to prevent a blow-up  
as $t\rightarrow 0$.

The relation (\ref{eq:linv}) has a simple interpretation. 
Since $H\sim 1/t$ and t  $\grad^2 \phig\sim -\delta$,  it  
 gives the intuitive relation $\vv \sim -\vnabla \phig  t$ between 
 the acceleration, $-\vnabla \phig$ and velocity. 
 The relation has played a prominent role in 
the analysis of large scale structure.
The density contrast $\delta(\vx)$ as  estimated  form the
distribution of galaxies, could be used in this relation to 
obtain the associated peculiar velocity $\vv(\vx)$.
This velocity fields could be compared with the 
actual observed velocities of galaxies. 
A good agreement between the fields yields the cosmological density parameter, $\Omega$, and also a confirmation of the gravitational 
instability mechanism for structure formation.
But,  perhaps more interestingly,  any mismatch between the fields
could be an indication of strange mode of galaxy/structure formation the result of which is a galaxy distribution different from that of the dark matter.

\subsection{Non-linear cosmological dynamics}
Linear theory is valid only when the fluctuations are small. 
In practice this is achieved by smoothing the observed 
galaxy distribution on small scales ($\lsim 10 \rm Mpc$). 
We describe here some non-linear methods which can be used for 
a variety of purposes, e.g. recovery of the initial conditions, estimating $\vv$ from the galaxy distribution and 
constraining the masses of galactic halos. 
Here we focus on the estimation of $\vv$. 
One can use numerical simulations of non-linear gravity to 
calibrate  semi-analytical non-linear 
 generalizations to (\ref{eq:linv}).
 The approach is useful as it provides partial differential equations  which can be solved for $\vv$ for a given source term, $\delta$. 
 Nevertheless, such generalizations are usually statistical 
 in nature.  
 In the following, we will describe a  more rigorous and accurate 
 approach.

We switch to a Lagrangian description for a system of $N$ equal mass 
particles in an expanding universe. Each particle represents a patch of matter  which, for practical purposes, could be a galaxy.  
The equations of motion are ($i=1\cdots N$),
\begin{equation}
\frac{\dd \vv_i}{\dd t}+2H\vv_i=\vg_i\; ,
\end{equation}
where   $\vg=-\vnabla\phig $ and is given by  
\begin{equation}
\vg(\vx)=-
\frac{G}{a^3}\sum_i \frac{\vx-\vx_i} {|\vx-\vx_i|^3} 
+\frac{4}{3}G\bar \rho a \vx \, , 
\label{eq:pot}
\end{equation}
  The equations  can be 
derived from  the action,
\begin{eqnarray}
\nonumber
{\mathrm S}&=&\int_0^{t_0} \dd t  \sum_i \\ 
&  &\left\{\frac{a^2}{2}  \vv_i^2 +
\frac{G}{a}  \left[   \sum_{j<i}\frac{1}{|\vx_i-\vx_j|}
+\frac{2\pi }{3} \bar \rho a^3\vx_i^2 \right]   \right\} \, 
\label{eq:lapd}
\end{eqnarray}
under stationary first variations of the orbits that leave $\vx $
fixed at the present epoch
and satisfy the constraint $t^{1/3}\vv\rightarrow const$ as
  $t\rightarrow 0$ \cite{Peebles1989,NusserBranchini,phelpsnusser}. The second condition on the
  velocities guarantees homogeneity near the big bang singularity $t \rightarrow 0$, preventing a blow up of the solutions. We
  expand the orbits in a time dependent base functions $\qn(t)$
  in
  the form,
\begin{equation}
\vx_i(t)=\vx_{i,0}+\sum_{n=1}^{n_{max}} \qn(t) \vC_{i,n}  ,
\label{eq:expand} 
\end{equation}
where $\vx_{i,0}$ is the position of the particle $i$ at the present epoch, and
the vectors $\vC_{i,n}$ are the expansion coefficients with respect to
which the action is varied, i.e., they satisfy $\pa {\mathrm S}/\pa
\vC_{i,n}=0$.  The base functions $\qn$ are chosen such that the boundary conditions are satisfied.

\begin{figure*}
 \includegraphics[scale=0.5,clip=true]{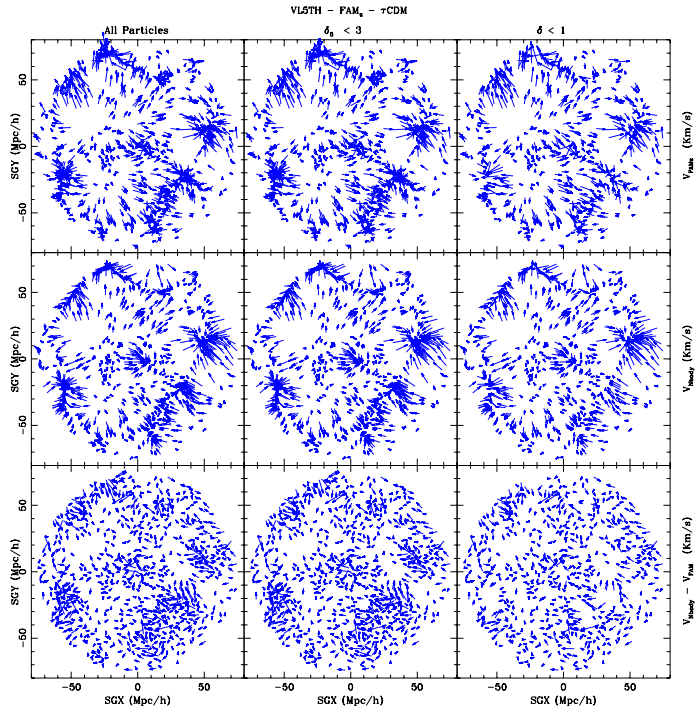}%
 \caption{\label{fig:fig}
 Maps of 2D-projected peculiar velocities for points residing
in a slice of thickness $6 \hmpc$ cut through a simulated 
catalog.
The length of the vectors is drawn in units of $1 \hmpc=50 \kms$.
The top row shows the least action predicted velocities (labeled FAMz). $N$-body
velocities are shown in the middle row. The velocity residuals, ${\bf
v}_{\scriptscriptstyle Nbody}-{\bf v}_{\scriptscriptstyle FAMz}$,
are displayed on the bottom.  The maps
shown in the panels to the left hand side refer to all the
points in the slice while only the velocities of points with moderate density contrast
are plotted in the central and right
columns. }
 \end{figure*}

Our strategy is to find orbits that are as close
as possible to the Hubble flow.  Therefore,  we search for the
minimum of the action and do not look for stationary points which might describe
oscillatory behavior of the orbits.  To find the
coefficients $\vC_{i,n}$ that
minimize the action, we  use the Conjugate Gradient Method (CGM)
which is fast and easy to implement.  
The gravitational force $\vg$ and its potential are computed using
the TREECODE gravity solver. The time
integration in the expression for the action is done using the
Gaussian quadrature method with  10 points at the time abscissa. 
The CGM requires an initial guess for $ \vC_{i,n} $.  
We will use the term FAM, for Fast Action Method, to refer to the 
reconstruction method described here. 
In the standard
FAM application we compute the initial guess using the 
linear theory relation between the velocity and mass distribution.
The minimum of the action proved to be rather insensitive to the
choice of initial guess for $ \vC_{i,n} $, as we have checked by
running  FAM experiments with initial $\vC_{i,n}$ both set
 to zero and  to random numbers with appropriate variance. 
Besides the initial set of $
\vC_{i,n} $, the other free parameters  are the softening used
by the gravity solver and the tolerance parameter that sets the
convergence of the CGM method. 
The success of the least action reconstruction method is illustrated in figure  (\ref{fig:fig}).

% Here is an example of the general form of a figure:
% Fill in the caption in the braces of the \caption{} command. Put the label
% that you will use with \ref{} command in the braces of the \label{} command.
% Use the figure* environment if the figure should span across the
% entire page instead of one column. There is no need to do explicit centering.

% If you have appendix consisting of several sections, uncomment 
% the following command. Use \appendix* if there is only one section 
% in the appendix
%\appendix
%\section{}

\section{Discussion}

The rapid rotation of galactic disks revealed the existence of 
dark matter halos which engulf the luminous component. 
The measured virial motions of galaxies in clusters of galaxies
also require the a gravitationally dominant dark component. 
Away from bound systems of galaxies and galaxy clusters, 
field galaxies show coherent flow pattern which deviates from 
a pure Hubble expansion. 
This coherent  velocity field is a direct probe of the large scale 
 dark matter distribution  
in as much as rotational speeds and virial motions are a measure of the 
dark matter in galaxies and clusters. 
 Indeed, the cosmic 
gravitational field responsible for the motions of galaxies, 
mainly  depends on the gravitationally dominant mass density field of the 
dark matter. 
The actual distribution of galaxies may well be quite different
from the dark matter distribution. 
Recent analysis of the galaxy surveys, however, reveal 
a good match between the statistical properties of the galaxy distribution 
and the corresponding properties for the dark matter as 
inferred from numerical simulations of dark matter evolution in the universe. 
This is encouraging, but there may still be significant deviations between the distribution of the dark and luminous components, which are not reflected 
in statistical comparisons. 
The only way to detect such deviations is via direct detailed comparisons 
between the measured velocities of galaxies and velocities estimated 
from the  galaxy distribution. These comparisons have been done in the linear regime. 
The overall  agreement between the fields is impressive, but 
minor persisting mismatch is detected in some  regions in the local volume. 
It is possible that non-linear analysis a al the least action principle 
could mitigate some of the disagreement. This remains to be seen. 
The least action principle could also be used to recover the initial conditions \cite{Mohayeehere},  allowing  us  to answer one of the fundamental question of whether or not  initial fluctuations
were gaussian \cite{ndy}.

%Perhaps one of the immediate challenges for reconstruction methods is the Local Group (LG) of galaxies. 
%The LG comprises about 40 galaxies within a distance of 5 Mpc from us, including our own Mily Way (MW) and 
%its   more massive neighbor, M31. 
%The action principle has been applied to this system. 

The program is not without flaws. 
Many physical effects need to be addressed in detail. 
Most pressing is incorporating the assembly (or merging) history of galaxies. 
Galaxies reside in dark matter halos which form in a hierarchical manner from small to large.
Thus our own Milky Way galaxy for example, is likely to have had  a major merging 
activity some 8 Gyr ago. All reconstruction methods assume that galaxies are point tracers 
of the mass density field and do not account for merging effects. 
%Another problem is that the mass distribution is probed by galaxies. 
%It is assumed the mass density field is directly related to the distribution of galaxies. 
%This assumption is sustained on large scales where the power spectrum of the galaxy 
%distribution matches very well the prediction of gravitational clustering of dark matter alone.
%However, on small scales, where nonlinear effects are important,  the relation between the mass and light 
%is highly uncertain. 

\begin{acknowledgments}
I wish to express my thanks to the organizers  of this 
exceptional conference. 
\end{acknowledgments}

% Uncomment the following line if you use BibTeX to produce 
% the reference section
% \bibliography{EE250sample}

\end{document}